\documentclass[conference,final]{IEEEtran}

\usepackage{graphicx}
\usepackage{amsmath}
\usepackage{amssymb}
\usepackage{color}
\usepackage{ifpdf}
\usepackage{float}
\usepackage[utf8]{inputenc}
\usepackage{multirow}
\usepackage{rotating}
\usepackage{subfigure}
\usepackage{setspace}
\usepackage{moresize}

\usepackage{caption} 
\usepackage{url}
\usepackage{booktabs}
\usepackage{listings}
\usepackage{paralist}
\usepackage{wrapfig}
\usepackage{multirow}
\usepackage{ifpdf}
\usepackage{xspace}
\usepackage{keyval}
\usepackage{color}

\definecolor{listinggray}{gray}{0.95}
\definecolor{darkgray}{gray}{0.7}
\definecolor{commentgreen}{rgb}{0, 0.4, 0}
\definecolor{darkblue}{rgb}{0, 0, 0.4}
\definecolor{middleblue}{rgb}{0, 0, 0.7}
\definecolor{darkred}{rgb}{0.4, 0, 0}
\definecolor{brown}{rgb}{0.5, 0.5, 0}

\usepackage[normalem]{ulem}
\makeatletter
\def\cyanuwave{\bgroup \markoverwith{\lower3.5\p@\hbox{\sixly \textcolor{cyan}{\char58}}}\ULon}
\def\reduwave{\bgroup \markoverwith{\lower3.5\p@\hbox{\sixly \textcolor{red}{\char58}}}\ULon}
\def\blueuwave{\bgroup \markoverwith{\lower3.5\p@\hbox{\sixly \textcolor{blue}{\char58}}}\ULon}
\font\sixly=lasy6 %
\makeatother

\newif\ifdraft
\draftfalse
\ifdraft
\usepackage{xcolor}
\definecolor{ocolor}{rgb}{1,0,0.4}
\newcommand{\note}[1]{ {\textcolor{ocolor} { (***Note: #1) }}}
\newcommand{\terminology}[1]{ {\textcolor{red} {(Terminology used: \textbf{#1}) }}}

\newcommand{\jhanote}[1]{ {\textcolor{red} { ***shantenu: #1 }}}
\newcommand{\alnote}[1]{ {\textcolor{blue} { ***andreL: #1 }}}
\newcommand{\ipnote}[1]{ {\textcolor{purple} { ***giannis: #1 }}}
\newcommand{\gnote}[1]{ {\cyanuwave{#1}}}
\definecolor{orange}{rgb}{1,.5,0}
\definecolor{dandelion}{cmyk}{0,0.29,0.84,0}
\newcommand{\cnote}[1]{ {\textcolor{magenta} { ***GC: #1 }}}
\else
\newcommand{\terminology}[1]{}
\newcommand{\alnote}[1]{}
\newcommand{\amnote}[1]{}
\newcommand{\gnote}[1]{}
\newcommand{\cnote}[1]{}
\newcommand{\jhanote}[1]{}
\newcommand{\note}[1]{}
\newcommand{\ipnote}[1]{}
\fi

\newcommand{\yarn}{YARN\xspace}
\newcommand{\rp}{RADICAL-Pilot\xspace}

\newcommand{\pilot}{Pilot\xspace}
\newcommand{\pilots}{Pilots\xspace}
\newcommand{\pilotjob}{Pilot-Job\xspace}
\newcommand{\pilotjobs}{Pilot-Jobs\xspace}
\newcommand{\pilotcompute}{Pilot-Compute\xspace}

\newcommand{\pilotdata}{Pilot-Data\xspace}

\newcommand{\computeunitdescription}{Compute-Unit Description\xspace}

\newcommand{\computeunit}{Compute-Unit\xspace}
\newcommand{\computeunits}{Compute-Units\xspace}

\newcommand{\cu}{CU\xspace}
\newcommand{\cus}{CUs\xspace}

\newcommand{\up}{\vspace*{-1em}}

\lstdefinestyle{myListing}{
  frame=single,
  backgroundcolor=\color{listinggray},
  language=C,
  basicstyle=\ttfamily \footnotesize,
  breakautoindent=true,
  breaklines=true
  tabsize=2,
  captionpos=b,
  aboveskip=0em,
  belowskip=-2em,
}

\lstdefinestyle{myPythonListing}{
  frame=single,
  backgroundcolor=\color{listinggray},
  language=Python,
  basicstyle=\ttfamily \scriptsize,
  breakautoindent=true,
  breaklines=true
  tabsize=2,
  captionpos=b,
}

\ifpdf
\DeclareGraphicsExtensions{.pdf, .jpg, .tif}
\else
\DeclareGraphicsExtensions{.ps,  .eps, .jpg}
\fi

\usepackage{listings}
\usepackage{paralist}
\lstnewenvironment{code}[1][]%
{
\noindent
\minipage{1.0 \linewidth}
\vspace{0.5\baselineskip}
\lstset{
    language=Python,
    frame=single,
    captionpos=b,
    stringstyle=\ttfamily,
    basicstyle=\scriptsize\ttfamily,
    showstringspaces=false,#1}
}
{\endminipage}

\defaultleftmargin{1em}{}{}{}
\begin{document}

\title{Hadoop on HPC: Integrating Hadoop and Pilot-based Dynamic Resource
  Management}

\author{
  Andre Luckow$^{1,2*}$, Ioannis Paraskevakos$^{1*}$, George Chantzialexiou$^{1}$, Shantenu Jha$^{1**}$\\
  \small{\emph{$^{1}$ Rutgers University, Piscataway, NJ 08854, USA}}\\
  \small{\emph{$^{2}$ School of Computing, Clemson University, Clemson, SC 29634, USA}}\\   \small{\emph{$^{(*)}$Contributed equally to this work}}\\
  \small{\emph{$^{**}$Contact Author: \texttt{shantenu.jha@rutgers.edu}}}\\
  \up\up 
 }

\date{}
\maketitle

\begin{abstract}
  High-performance computing platforms such as ``supercomputers'' have
  traditionally been designed to meet the compute demands of scientific
  applications. Consequently, they have been architected as net producers and
  not consumers of data.  The Apache Hadoop ecosystem has evolved to meet the
  requirements of data processing applications and has addressed many of the
  traditional limitations of HPC platforms.  There exist a class of scientific
  applications however, that need the collective capabilities of traditional
  high-performance computing environments and the Apache Hadoop ecosystem.  For
  example, the scientific domains of bio-molecular dynamics, genomics and
  network science need to couple traditional computing with Hadoop/Spark based
  analysis.  We investigate the critical question of how to present the
  capabilities of both computing environments to such scientific
  applications. Whereas this questions needs answers at multiple levels, we
  focus on the design of resource management middleware that might support the
  needs of both.  We propose extensions to the Pilot-Abstraction so as to
  provide a unifying resource management layer. This is an important step
  towards interoperable use of HPC and Hadoop/Spark.  It also allows
  applications to integrate HPC stages (e.\,g.\ simulations) to data analytics.
  Many supercomputing centers have started to officially support Hadoop
  environments, either in a dedicated environment or in hybrid deployments using
  tools such as myHadoop. This typically involves many intrinsic,
  environment-specific details that need to be mastered, and often swamp
  conceptual issues like: How best to couple HPC and Hadoop application stages?
  How to explore runtime trade-offs (data localities vs. data movement)?  This
  paper provides both conceptual understanding and practical solutions to the
  integrated use of HPC and Hadoop environments.  Our experiments are performed
  on state-of-the-art production HPC environments and provide middleware for
  multiple domain sciences.

\end{abstract}

\section{Introduction}

The MapReduce~\cite{mapreduce} abstraction popularized by Apache
Hadoop~\cite{hadoop} has been successfully used for many data-intensive
applications in different domains~\cite{37684}.  One important differentiator of
Hadoop compared to HPC is the availability of many higher-level abstractions and
tools for data storage, transformations and advanced analytics. These
abstraction typically allow high-level reasoning about data parallelism without
the need to manually partition data, manage tasks processing this data and
collecting the results, which is required in other environments.  Within the
Hadoop ecosystem, tools like Spark~\cite{Zaharia:2010:SCC:1863103.1863113} have
gained popularity by supporting specific data processing and analytics needs and
are increasingly used in sciences, e.\,g.\ for DNA
sequencing~\cite{Massie:EECS-2013-207}.

Data-intensive applications are associated with a wide variety of
characteristics and properties, as summarized by Fox
et\,al.~\cite{bigdata-ogres,bigdata-use-cases-nist}.  Their complexity and
characteristics are fairly distinct from HPC applications. For example, they
often comprise of multiple stages such as, data ingest, pre-processing,
feature-extraction and advanced analytics. While some of these stages are I/O
bound, often with different patterns (random/sequential access), other stages
are compute-/memory-bound.  Not surprisingly, a diverse set of tools for data
processing (e.\,g.\ MapReduce, Spark RDDs), access to data sources (streaming,
filesystems) and data formats (scientific data formats (HDF5), columnar formats
(ORC and Parquet)) have emerged and they often need to be combined in order to
support the end-to-end needs of applications.

Some applications however, defy easy classification as data-intensive or HPC. In
fact, there is specific interest in a class of of scientific applications, such
as bio-molecular dynamics~\cite{doi:10.1146/annurev-biophys-042910-155245}, that
have strong characteristics of both data-intensive and HPC. Bio-molecular
simulations are now high-performant, reach increasing time scales and problem
sizes, and thus generating immense amounts of data.  The bulk of the data in
such simulations is typically trajectory data that is time-ordered set of
coordinates and velocity. Secondary data includes other physical parameters
including different energy components. Often times the data generated needs to
be analyzed so as to determine the next set of simulation configurations. The
type of analysis varies from computing the higher order moments, to principal
components, to time-dependent variations.

MDAnalysis~\cite{mdanalysis} and CPPTraj~\cite{doi:10.1021/ct400341p} are two
tools that evolved to meet the increasing analytics demands of molecular
dynamics applications; Ref~\cite{himach} represents an attempt to provide
MapReduce based solutions in HPC environments.  These tools provide powerful
domain-specific analytics; a challenge is the need to scale them to high data
volumes produced by molecular simulations as well as the coupling between the
simulation and analytics parts.  This points to the need for environments that
support scalable data processing while preserving the ability to run simulations
at the scale so as to generate the data.  To the best of our knowledge, there
does not exist a solution that provides the integrated capabilities of Hadoop
and HPC.  For example, Cray's analytics platform
Urika~\footnote{http://www.cray.com/products/analytics} has Hadoop and Spark
running on HPC architecture as opposed to regular clusters, but without the HPC
software environment and capabilities.  However, several applications ranging
from bio-molecular simulations to epidemiology models~\cite{network1} require
significant simulations interwoven with analysis capabilities such as clustering
and graph analytics; in other words some stages (or parts of the same stage) of
an application would ideally utilize Hadoop/Spark environments and other stages
(or parts thereof) utilize HPC environments.

Over the past decades, the High Performance Distributed Computing (HPDC)
community has made significant advances in addressing resource and workload
management on heterogeneous resources. For example, the concept of multi-level
scheduling~\cite{1392910} as manifested in the decoupling of workload assignment
from resource management using the concept of intermediate container jobs (also
referred to as \pilotjobs~\cite{pstar12}) has been adopted for both HPC and
Hadoop.  Multi-level scheduling is a critical capability for data-intensive
applications as often only application-level schedulers can be aware of the
localities of the data sources used by a specific application.  This motivated
the extension of the \pilot-Abstraction to
\pilotdata~\cite{pilot-data-jpdc-2014} to form the central component of
a resource management middleware.

In this paper, we explore the integration between Hadoop and HPC resources
utilizing the \pilot-Abstraction allowing application to manage HPC (e.\,g.\
simulations) and data-intensive application stages in a uniform way. We propose
two extensions to \rp: the ability to spawn and manage Hadoop/Spark clusters on
HPC infrastructures on demand (Mode I), and to connect and utilize Hadoop and
Spark clusters for HPC applications (Mode II).  Both extensions facilitate the
complex application and resource management requirements of data-intensive
applications that are best met by a best-of-bread mix of Hadoop and HPC. By
supporting these two usage modes, \rp dramatically simplifies the barrier of
deploying and executing HPC and Hadoop/Spark side-by-side.

This paper is structured as follows: In Section~\ref{sec:related} we
survey the Hadoop ecosystem and compare it to HPC. We continue with a discussion of the 
new \pilot-based capabilities that support
Hadoop/HPC interoperability in section~\ref{sec:pilot-data-hadoop}. The results of our
experimental validation are presented in section~\ref{sec:experiments}. We
conclude with a discussion of the contributions and lessons learnt, as well as
relevant future work in section~\ref{sec:conclusion}.

\section{Background and Related Work}
\label{sec:related}

Our approach is to design a common software environment while attempting to be
agnostic of specific hardware infrastructures, technologies and trends. In this section, 
we provide background information and comparative information on system
abstractions, resource management and  interoperability in HPC and Hadoop.

Hadoop~\cite{hadoop} has evolved to become the standard implementation of the
MapReduce abstraction on top of the Hadoop filesystem and the \yarn resource
management.  In fact, over the past years, Hadoop evolved to a general purpose
cluster computing framework suited for data-intensive applications in
industry~\cite{luckow-2015} and sciences~\cite{tale-of-two-ieee-2014}.

HPC and Hadoop originated from the need to support different kinds of applications:
compute-intensive applications in the case of HPC, and data-intensive in the case of
Hadoop. Not surprisingly, they follow different design paradigms: In HPC environments,
storage and compute are connected by a high-end network (e.\,g.\ Infiniband) with
capabilities such as RDMA; Hadoop co-locates both. HPC infrastructures introduced
parallel filesystems, such as Lustre, PVFS or GPFS, to meet the increased I/O demands of
data-intensive applications and archival storage and to address the need for retaining
large volumes of primary simulation output data. The parallel filesystem model of using
large, optimized storage clusters exposing a POSIX compliant rich interface and
connecting it to compute nodes via fast interconnects works well for compute-bound task.
It has however, some limitations for data-intensive, I/O-bound workloads that require a
high sequential read/write performance. Various approaches for integrating parallel
filesystems, such as Lustre and PVFS, with Hadoop
emerged~\cite{lustre_hadoop,Tantisiriroj:2011:DDF:2063384.2063474}, which yielded good
results in particular for medium-sized workloads.

While Hadoop simplified the processing of vast volumes of data, it has
limitations in its expressiveness as pointed out by various
authors~\cite{magalan,dryad}.  The complexity of creating sophisticated
applications such as iterative machine learning algorithms required multiple
MapReduce jobs and persistence to HDFS after each iteration.  This is lead to
several higher-level abstractions for implementing sophisticated data pipelines.
Examples of such higher-level execution management frameworks for Hadoop are:
Spark~\cite{Zaharia:2010:SCC:1863103.1863113}, Apache Flink~\cite{flink}, Apache
Crunch~\cite{crunch} and Cascading~\cite{cascading}.

The most well-known emerging processing framework in the Hadoop ecosystem is
Spark~\cite{Zaharia:2010:SCC:1863103.1863113}. In contrast to 
MapReduce, it provides a richer API, more language bindings and a
novel memory-centric processing engines, that can utilize distributed memory and
can retain resources across multiple task generation. Spark's \emph{Reliable
  Distributed Dataset (RDD)} abstraction provides a powerful way to manipulate
distributed collection stored in the memory of the cluster nodes. Spark is
increasingly used for building complex data workflows and advanced analytic
tools, such as MLLib~\cite{mllib} and SparkR.

Although the addition/development of new and higher-level execution frameworks
addressed some of the problems of data processing, it introduced the problem of
heterogeneity of access and resource management, which we now discuss.

Hadoop originally provided a rudimentary resource management system, the \yarn
scheduler~\cite{yarn-paper} provides a robust application-level
scheduler framework addressing the increased requirements with respect to applications 
and infrastructure: more complex data localities (memory, SSDs, disk, rack, datacenter),
long-lived services, periodic jobs, interactive and batch jobs need to be
supported on the same environment. In contrast, to traditional batch schedulers,
\yarn is optimized for data-intensive environments supporting data-locality and
the management of a large number of fine-granular tasks (found in data-parallel
applications). 

While \yarn manages system-level resources, applications and runtimes have to implement
an application-level scheduler that optimizes their specific resource requirements,
e.\,g.\ with respect to data locality. This application-level scheduler is referred to as
{\it Application Master} and is responsible for allocating resources -- the so called
containers -- for the applications and to execute tasks in these containers. Data
locality, e.\,g.\ between HDFS blocks and container locations, need to managed by the
Application Master by requesting containers on specific nodes/racks.

Managing resources on top of \yarn is associated with several challenge: while
fine-grained, short-running tasks as found in data-parallel MapReduce applications are
well supported, other workload characteristics are less well supported, e.\,g.\
gang-scheduled parallel MPI applications and long-running applications found
predominantly in HPC environments. To achieve interoperability and integration between 
Hadoop and HPC, it is essential to consider a more diverse set of workloads on top of 
YARN.

To achieve interoperability, several frameworks explore the usage of Hadoop on HPC
resources. Various frameworks for running Hadoop on HPC emerged, e.\,g., Hadoop on
Demand~\cite{hod}, JUMMP~\cite{6702650}, MagPie~\cite{magpie},
MyHadoop~\cite{Krishnan04myhadoop}, MyCray~\cite{mycray}. While these frameworks can
spawn and manage Hadoop clusters many challenges with respect to optimizing
configurations and resource usage including the use of available SSDs for the shuffle
phase, of parallel filesystems and of high-end network features, e.\,g.\ RDMA~\cite{homr}
remain. Further, these approaches do not address the need for interoperability between
HPC and Hadoop application stages.

A particular challenge for Hadoop on HPC deployment is the choice of storage and
filesystem backend. Typically, for Hadoop local storage is preferred; nevertheless, in
some cases, e.\,g.\ if many small files need to processed, random data access is required
or the number of parallel tasks is low to medium, the usage of Lustre or another parallel
filesystem can yield in a better performance. For this purpose, many parallel filesystems
provide a special client library, which improves the interoperability with Hadoop; it
limits however data locality and the ability for the application to optimize for data
placements since applications are commonly not aware of the complex storage hierarchy.
Another interesting usage mode is the use of Hadoop as active archival storage -- in
particular, the newly added HDFS heterogeneous storage support is suitable for supporting
this use case.

Another challenge is the integration between both HPC and Hadoop environments.
Rather than preserving HPC and Hadoop ``environments'' as software silos,
there is a need for an approach that integrates them.  We propose the
\pilot-Abstraction as unifying concept to efficiently support the integration,
and not just the interoperability between HPC and Hadoop. By utilizing the
multi-level scheduling capabilities of \yarn, \pilot-Abstraction can efficiently
manage Hadoop cluster resources providing the application with the necessary
means to reason about data and compute resources and allocation. On the other
side, we show, how the \pilot-Abstraction can be used to manage Hadoop applications
on HPC environments.

The \pilot-Abstraction~\cite{pstar12} has been successfully used in HPC
environments for supporting a diverse set of task-based workloads on distributed
resources. A \pilotjob is a placeholder job that is submitted to the resource
management system representing a container for a dynamically determined set of
compute tasks. \pilotjobs are a well-known example of multi-level scheduling,
which is often used to separate system-level resource and user-level workload
management. The \pilot-Abstraction defines the following entities: A
\pilotcompute allocates a set of computational resources (e.\,g.\,cores); a
\computeunit (CU) as a self-contained piece of work represented as executable
that is submitted to the \pilotjob. A \cu can have data dependencies, i.\,e.\ a
set of files that need to be available when executing the \cu. A workloads
typically consists of a set of dependents \cus. The \pilot-Abstraction has been
implemented within BigJob~\cite{pstar12,saga_bigjob_condor_cloud} and its second
generation prototype \rp~\cite{review_radicalpilot_2015}. The interoperability
layer of both frameworks is SAGA~\cite{saga-x}, which is used for accessing the
resource management system (e.\,g.\ SLURM, Torque and SGE) and for file
transfers. SAGA is a lightweight interface that provides standards-based
interoperable capabilities to the most commonly used functionalities required to
develop distributed applications, tools and services.

\section{Integrating Hadoop and Spark with RADICAL-Pilot}
\label{sec:pilot-data-hadoop}

An important motivation of our work is to provide advanced and scalable data
analysis capabilities for existing high-performance applications (e.g.,
large-scale molecular dynamics simulations). This requires adding data-intensive
analysis while preserving high-performance computing capabilities. Having
established the potential of the \pilot-Abstraction for a range of
high-performance
applications~\cite{review_repex_2016,ethread-biomed-2014,cfd-multiscale-2014},
we use it as the starting point for integrated high-performance compute and
data-intensive analysis.  We propose several extensions to \rp to facilitate the
integrated use of HPC and Hadoop frameworks using the \pilot-Abstraction.

\begin{figure}[t]
    \centering
    \includegraphics[width=.45\textwidth]{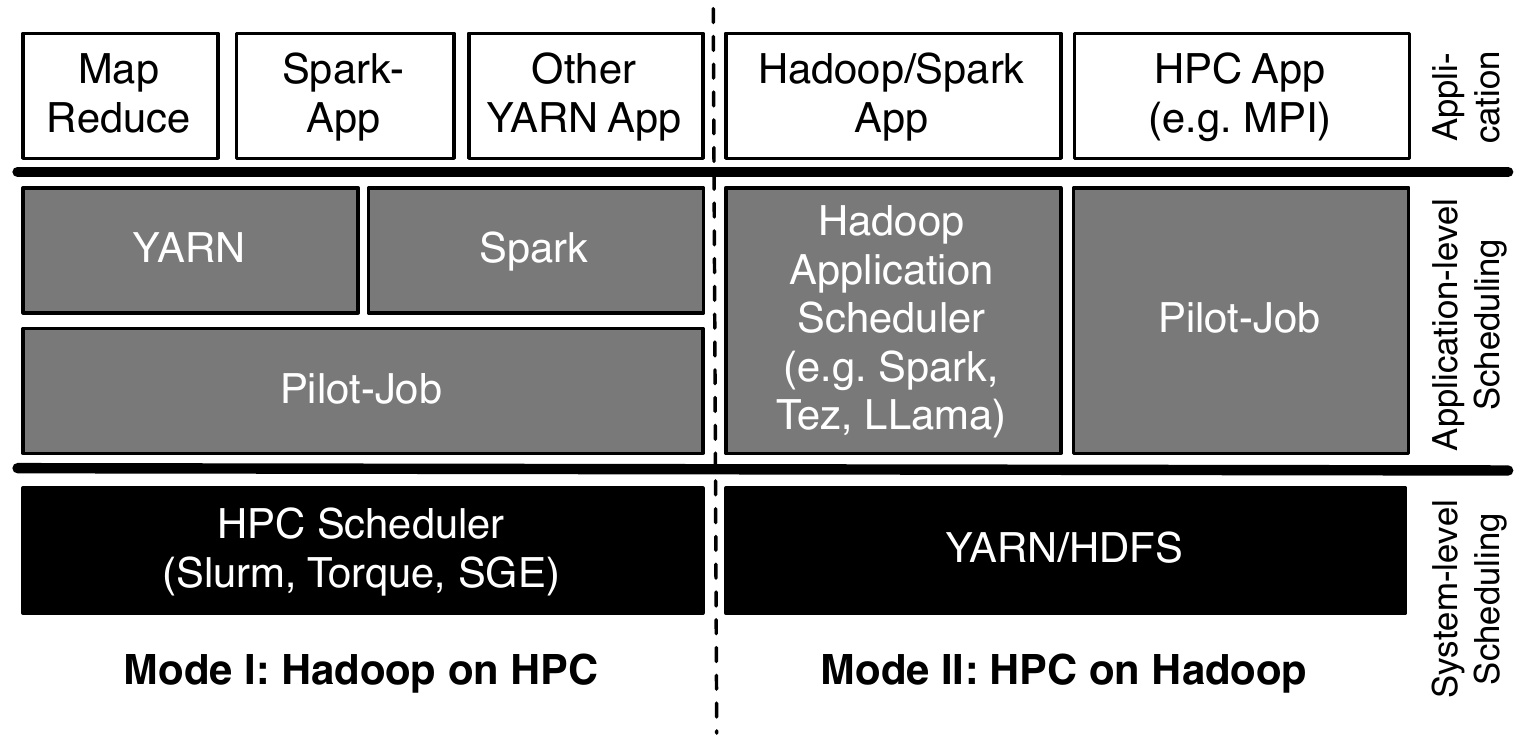}
    \caption{\textbf{Hadoop and HPC Interoperability:} There are two usage modes: 
    (i) spawning a \yarn or Spark cluster on a HPC environment (Hadoop on HPC), and (ii) Running 
    HPC applications inside a \yarn cluster (HPC on Hadoop).  
\label{fig:figures_hadoop-on-hpc-viceverse}}
\end{figure}

As depicted in Figure~\ref{fig:figures_hadoop-on-hpc-viceverse}, there are at
least two different usage modes to consider:
\begin{compactenum}[(i)]
\item Mode I: Running Hadoop/Spark applications on HPC environments (Hadoop on
  HPC),
\item Mode II: Running HPC applications on YARN clusters (HPC on Hadoop).
\end{compactenum}

Mode I is critical to support traditional HPC environments (e.\,g., the majority
of XSEDE resources) so as to support applications with both compute and data
requirements.  Mode II is important for cloud environments (e.\,g.\ Amazon's
Elastic MapReduce, Microsoft's HDInsight) and an emerging class of HPC machines
with new architectures and usage modes, such as Wrangler~\cite{wrangler} that
support Hadoop natively. For example, Wrangler supports dedicated Hadoop
environments (based on Cloudera Hadoop 5.3) via a reservation mechanism.

In the following we propose a set of tools for supporting both of these usage
modes: In section~\ref{sec:saga_hadoop} we present SAGA-Hadoop, a light-weight,
easy-to-use tool for running Hadoop on HPC (Mode I). We then discuss, the
integration of Hadoop and Spark runtimes into \rp, which enables both the
interoperable use of HPC and Hadoop, as well as the integration of HPC and
Hadoop applications (Mode I and II) (Section~\ref{sec:pilot_hadoop_hpc}
to~\ref{sec:rp_spark}). Using these new capabilities, applications can
seamlessly connect HPC stages (e.\,g.\ simulation stages) with analysis staging
using the \pilot-Abstraction to provide unified resource management.

\subsection{SAGA-Hadoop: Supporting Hadoop/Spark on HPC}
\label{sec:saga_hadoop}

SAGA-Hadoop~\cite{saga-hadoop} is a tool for supporting the deployment of Hadoop
and Spark on HPC resources (Mode I in
Figure~\ref{fig:figures_hadoop-on-hpc-viceverse}). Using SAGA-Hadoop an
applications written for \yarn (e.\,g.\ MapReduce) or Spark (e.\,g. PySpark, DataFrame and MLlib 
applications) can be executed on HPC resources.

\begin{figure}[t]
    \centering
\includegraphics[width=.35\textwidth]{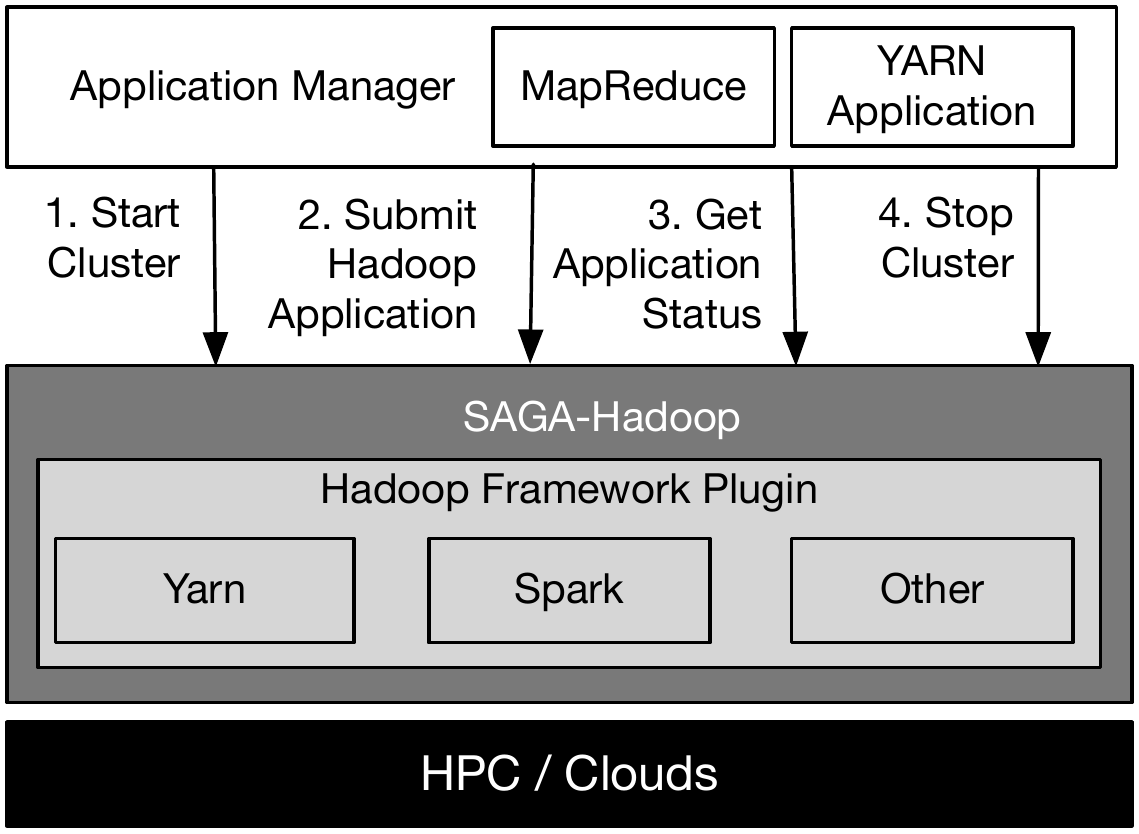}
    \caption{\textbf{SAGA-Hadoop for HPC and Cloud Infrastructures:} 
    SAGA-Hadoop provides  uniform framework to managing Hadoop and Spark 
    clusters on resources managed by HPC schedulers, such as PBS, SGE and 
    SLURM. }
    \label{fig:saga-hadoop}
\end{figure}

Figure~\ref{fig:saga-hadoop} illustrates the architecture of
SAGA-Hadoop. SAGA-Hadoop uses SAGA~\cite{saga-x} to spawn and control Hadoop
clusters inside an environment managed by an HPC scheduler, such as PBS, SLURM
or SGE, or clouds. SAGA is used for dispatching a bootstrap process that
generates the necessary configuration files and for starting the Hadoop
processes. The specifics of the Hadoop framework (i.\,e.\ \yarn and Spark) are
encapsulated in a Hadoop framework plugin (also commonly referred to as
adaptors). SAGA-Hadoop delegates tasks, such as the download, configuration and
start of a framework to this plugin. In the case of YARN, the plugin is then
responsible for launching \yarn's Resource and Node Manager processes; in the
case of Spark, the Master and Worker processes. This architecture allows for
extensibility -- new frameworks, e.\,g.\ Flink, can easily be added.

While nearly all Hadoop frameworks (e.\,g.\ MapReduce and Spark) support \yarn
for resource management, Spark provides a standalone cluster mode, which is more
efficient in particular on dedicated resources. Thus, a special adaptor for
Spark is provided. Once the cluster is setup, users can submit applications
using a simple API that allows them to start and manage YARN or Spark
application processes. 

While SAGA-Hadoop provides the interoperability between YARN and HPC resources
by treating YARN as a substitute for SLURM or Torque, the integration of YARN
and HPC application or application stages remains challenging. In the following,
we explore the usage of the \pilot-Abstraction, an implementation of which is
\rp, to enable the integration between these different application types.

\subsection{\rp and YARN Overview}
\label{sec:pilot_hadoop_hpc}
\label{sec:rp-impl}

With the introduction of \yarn, a broader set of applications can be executed
within Hadoop clusters than earlier.  However, developing and deploying YARN
applications potentially side-by-side with HPC applications remains a difficult
task. Established abstractions that are easy-to-use while enabling the user to
reason about compute and data resources across infrastructure types (i.\,e.\
Hadoop, HPC and clouds) are missing. \jhanote{we should be consistent with
  ``infrastructure'' (e.g. Wrangler, Stampede) versus ``infrastructure type''
  (e.g., Hadoop, Cloud etc).}

Schedulers such as \yarn effectively facilitate application-level scheduling,
the development efforts for \yarn applications are high. \yarn provides a
low-level abstraction for resource management, e.g., a Java API and protocol
buffer specification. Typically interactions between YARN and the applications
are much more complex than the interactions between an application and a HPC
scheduler. Further, applications must be able to run on a dynamic set of
resources; \yarn e.\,g.\ can preempt containers in high-load situations.
Data/compute locality need to be manually managed by the application scheduler
by requesting resources at the location of an file chunk. Also, allocated
resources (the so called \yarn containers) can be preempted by the scheduler.

To address these shortcomings, various frameworks that aid the development of
YARN applications have been proposed:
Llama~\cite{llama} offers a long-running application master for \yarn designed
for the Impala SQL engine. Apache Slider~\cite{apache-slider} supports
long-running distributed application on \yarn with dynamic resource needs
allowing applications to scale to additional containers on demand.  While these
frameworks simplify development, they do not address concerns such as
interoperability and integration of HPC/Hadoop. In the following, we explore
the integration of YARN into the \rp (RP) framework. This approach allows
applications to run HPC and YARN application parts side-by-side.

\begin{figure*}
\centering
\includegraphics[width=0.8\textwidth]{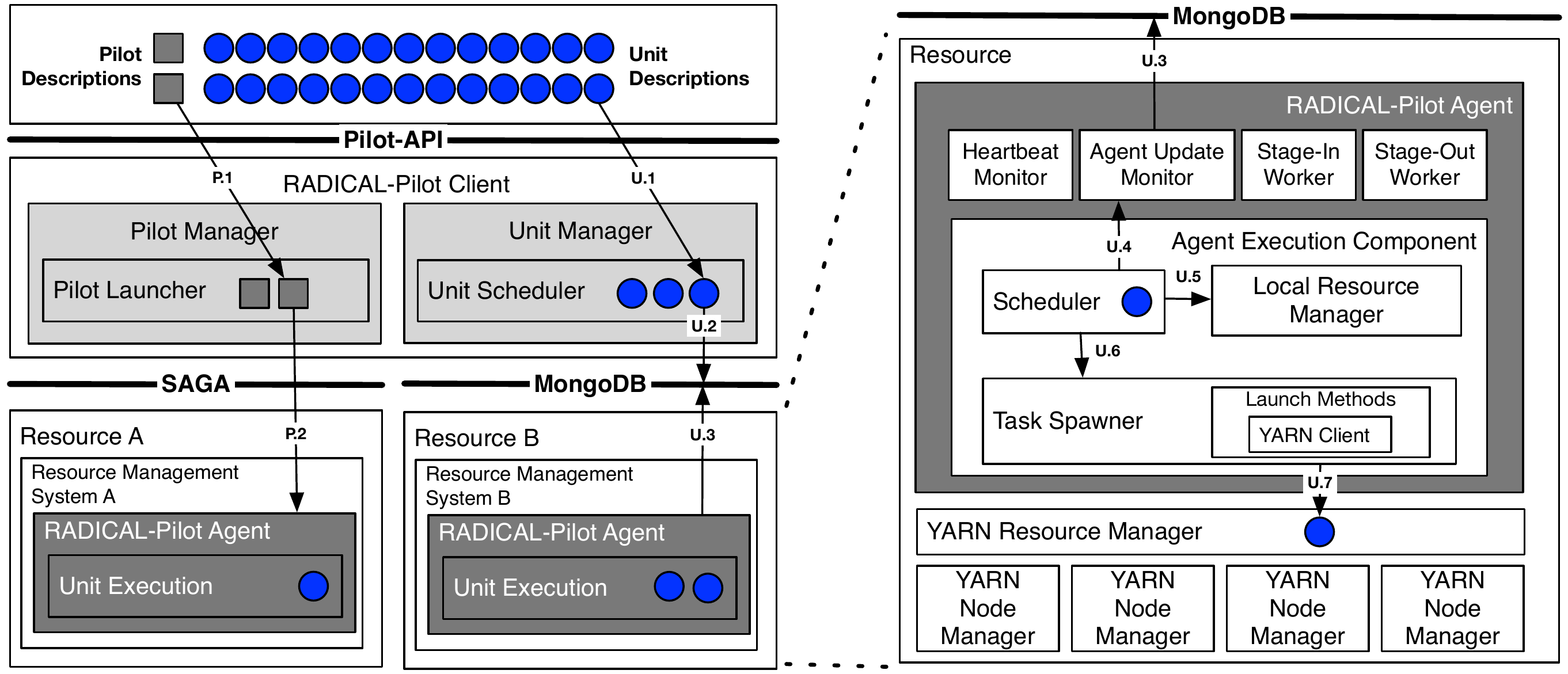}
\caption{\textbf{RADICAL-Pilot and YARN Integration:} There are two main interactions between the 
application and \rp -- the management of \pilots (P.1-P.2) and the management of \computeunits 
(U.1-U.7). All YARN specifics are encapsulated in the RADICAL-Pilot-Agent. \label{fig:comp_rp_arch} 
}
\end{figure*}

Figure~\ref{fig:comp_rp_arch} illustrates the architecture of \rp and the
components that were extended for YARN.  The figure on the left shows the macro
architecture of \rp; the figure on the right a blown-up look into the
architecture of the Pilot-Agent which is a critical functional component.  \rp
consists of a client module with the \pilot-Manager and Unit-Manager and a set
of \rp-Agents running on the resource.  The Pilot-Manager is the central entity
responsible for managing the lifecycle of a set of Pilots: \pilots are described
using a \pilot description, which contains the resource requirements of the
Pilot and is submitted to the \pilot-Manager. The \pilot-Manager submits the
placeholder job that will run the \rp-Agent via the Resource Management System
using the SAGA-API (steps P.1-P.7). Subsequently, the application workload (the
\computeunits) is managed by the Unit-Manager and the \rp-Agent (steps
U.1-U.7). More details are available at~\cite{review_radicalpilot_2015}.

The \rp-Agent has a modular and extensible architecture and consists of the
following components: the Agent Execution Component, the Heartbeat Monitor,
Agent Update Monitor, Stage In and Stage Out Workers. The main integration of
YARN and RP is done in the Agent Execution Component. This component consist of
four sub-components: The {\bf scheduler} is responsible for monitoring the
resource usage and assigning CPUs to a \computeunit. The {\bf Local Resource
  Manager} (LRM) interacts with the batch system and communicates to the Pilot
and Unit Managers the number of cores it has available and how they are
distributed. The {\bf Task Spawner} configures the execution environment,
executes and monitors each unit. The {\bf Launch Method} encapsulates the
environment specifics for executing an applications, e.\,g.\ the usage of
\texttt{mpiexec} for MPI applications, machine-specific launch methods (e.g.
\texttt{aprun} on Cray machines) or the usage of YARN. After the Task Spawner
completes the execution of a unit, it collects the exit code, standard input and
output, and instructs the scheduler about the freed cores.

\subsection{Integration of \rp and YARN}

There are two integration options for \rp and YARN: (i) Integration on
\pilot-Manager level, via a SAGA adaptor, and (ii) integration on the \rp-Agent
level. The first approach is associated with several challenges: firewalls
typically prevent the communication between external machines and a YARN
clusters. A YARN application is not only required to communicate with the
resource manager, but also with the node managers and containers; further, this
approach would require significant extension to the \pilot-Manager, which
currently relies on the SAGA Job API for launching and managing \pilots. 
Capabilities like the on-demand provisioning of a YARN cluster and
the complex application-resource management protocol required by YARN are
difficult to abstract behind the SAGA API.  

The second approach encapsulated YARN specifics on resource-level. If required,
a YARN cluster is de-centrally provisioned. Units are scheduled and submitted to
the YARN cluster via the Unit-Manager, the MongoDB-based communication protocol
and the \rp-Agent scheduler.  By integrating at the \rp-Agent level, \rp
supports both Mode I and II as outlined in
Figure~\ref{fig:figures_hadoop-on-hpc-viceverse}.

As illustrated in Figure~\ref{fig:comp_rp_arch}, in the first phase (step P.1
and P.2) the \rp-Agent is started on the remote resource using SAGA, e.\,g.\
SLURM. In Mode I, during the launch of the \rp-Agent the YARN cluster is spawned
on the allocated resources (Hadoop on HPC); in Mode II the \rp-Agent will just
connect to a YARN cluster running on the machine of the \rp-Agent.  Once the
\rp-Agent has been started, it is ready to accept \computeunits submitted via
the Unit-Manager (step U.1). The Unit-Manager queues new \computeunits using a
shared MongoDB instance (step U.2). The \rp-Agent periodically checks for new
\computeunits (U.3) and queues them inside the scheduler (U.4). The execution of
the \computeunit is managed by the Task Spawner (step U.6 and U.7). In the
following, we describe how these components have been extended to support \yarn.

\jhanote{Maybe a discussion of components (where applicable) can be broken up
  into two parts: what the component does and how it was extended. Currently
  reads a bit difficult.}\alnote{done}

The \emph{Local Resource Manager (LRM)} provides an abstraction to local resource details for other
components of the \rp-Agent. The LRM evaluates the environment variables provided by the 
resource management systems to obtain information, such as the number of cores per node, memory and 
the assigned nodes.  This information can be accessed through the Resource Manager's REST API. As 
described, there are two deployment modes. In Mode I (Hadoop on HPC), during the
initialization of the \rp-Agent, the LRM setups the HDFS and \yarn daemons: First, the LRM 
downloads Hadoop and creates the necessary configuration files, i.\,e. the 
\texttt{mapred-site.xml}, \texttt{core-site.xml}, \texttt{hdfs-site.xml}, \texttt{yarn-site.xml} 
and the slaves and master file containing the allocated nodes. The node that is running the Agent 
are assigned to run the master daemons: the HDFS Namenode and the \yarn Resource Manager. After 
the configuration files are written, HDFS and YARN are started and meta-data about the cluster,
i.\,e.\ the number of cores and memory, are provided to the scheduler. They remain active until all
the tasks are executed. Before termination of the agent, the LRM stops the Hadoop and \yarn daemons
and removes the associated data files. In Mode II (Hadoop on HPC), the LRM solely collects the 
cluster resource information.

The \emph{scheduler} is another extensible component of the \rp-Agent responsible for queueing
compute units and assigning these to resources. For YARN we utilize a special scheduler that
utilizes updated cluster state information (e.\,g.\ the amount of available virtual cores, memory,
queue information, application quotas etc.) obtained via the Resource Manager's REST API. In 
contrast to other \rp schedulers, it specifically utilizes memory in addition to cores for 
assigning resource slots.

\emph{Task Spawner and Launch Method:} The Task Spawner is responsible for managing and monitoring
the execution of a \computeunit. The Launch Method components encapsulates resource/launch-method
specific operations, e.\,g.\ the usage of the \texttt{yarn} command line tool for submitting and
monitoring applications. After launch of a \computeunit, the Task Spawner periodically monitors 
its execution and updates its state in the shared MongoDB instance. For YARN the application log 
file is used for this purpose.

\begin{figure}[t]
  \centering
    \includegraphics[width=.40\textwidth]{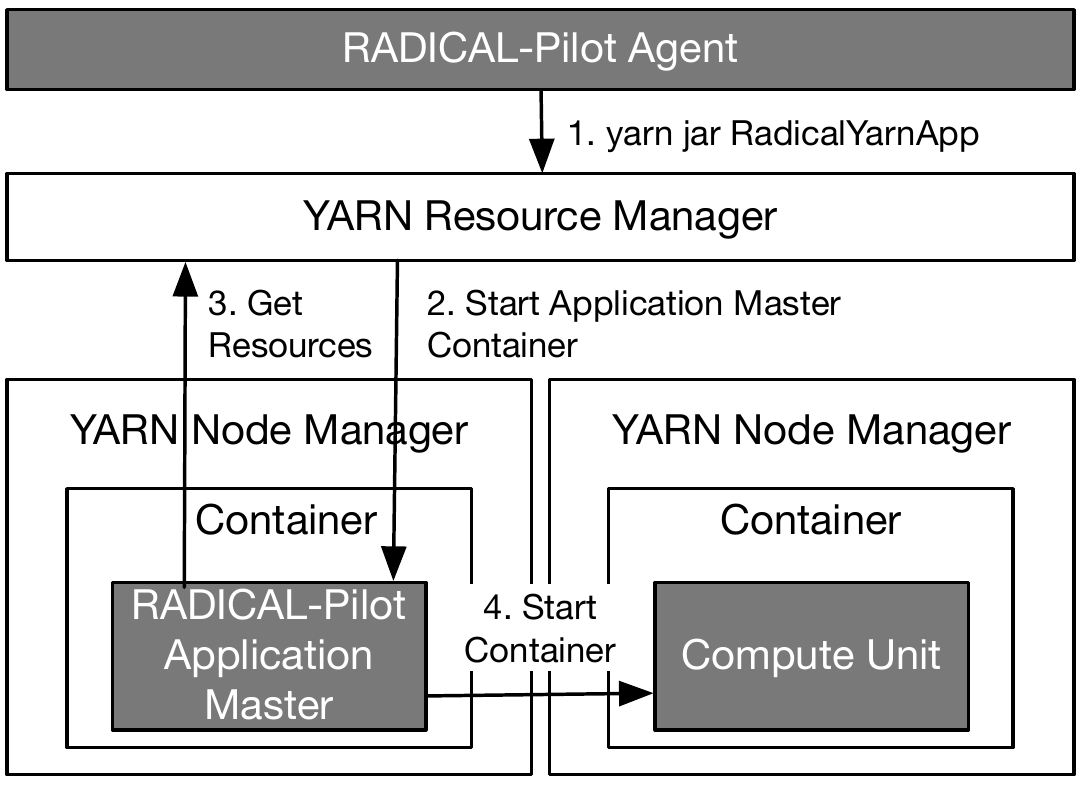}
    \caption{\textbf{\rp YARN Agent Application: } \rp provides a YARN application that manages the 
    execution of \computeunits. The application is initialized with parameters defined in the 
    \computeunitdescription and started by the Task Spawner (step 1/2). The Application Master 
    requests resources from the Resource Manager and starts a container running the 
    \computeunit (step 3/4).}
  \label{fig:figures_yarn}
\end{figure}

\emph{\rp Application Master:} A particular integration challenge is the multi-step resource allocation process imposed by \yarn depicted in Figure~\ref{fig:figures_yarn}, which differs
significantly from HPC schedulers. The central component of a YARN application
is the Application Master, which is responsible for negotiating resources with
the YARN Resource Manager as well as for managing the execution of the
application in the assigned resources. The unit of allocation in YARN is a so
called container (see~\cite{Murthy:2014:AHY:2636998}). The \yarn client (part of
the YARN Launch Method) implements a so-called \yarn Application Master, which
is the central instance for managing the resource demands of the
application. \rp utilizes a managed application master that is run inside a YARN
container. Once the Application Master container is started, it is responsible
for subsequent resource requests; in the next step it will request the a YARN
container meeting the resource requirements of the \computeunit from the
Resource Manager.  Once a container is allocated by \yarn, the \cu will be
started inside these containers. A wrapper script responsible for setting up a
\rp environment, staging of the specified files and running the executable
defined in the \computeunitdescription is used for this purpose. Every
\computeunit is mapped to a YARN application consisting of an Application Master
and a container of the size specified in the \computeunitdescription. In the
future, we will further optimize the implementation by providing support for
Application Master and container re-use.

\subsection{Spark Integration}
\label{sec:rp_spark}

Spark offers multiple deployment modes: standalone, YARN and Mesos. While it is possible to
support Spark on top of YARN, this approach is associated with significant complexity and overhead
as two instead of one framework need to be configured and run. Since we \rp operates in user-space
and single-user mode, no advantages with respect to using a multi-tenant YARN cluster environment
exit. Thus, we decided to support Spark via the standalone deployment mode.

Similar to the YARN integration, the necessary changes for Spark are confined to the 
\rp-Agent. Similarly, the Local Resource Manager is mainly responsible for initialization and 
deployment of the Apache Spark environment. In the first step the LRM detects the number of cores,  
memory and nodes provided by the Resource Management System, verifies and 
downloads necessary dependencies (e.\,g.\ Java, Scala, and the necessary Spark binaries). It then 
creates the necessary configuration files, e.\,g.\ \texttt{spark-env.sh},  \texttt{slaves} and 
\texttt{master} files,  required for running a multi-node, standalone Spark 
cluster.
Finally, the LRM is starting the Spark cluster using the previously generated configuration. 
Similar to YARN, a Spark \rp-Agent scheduler is used for managing Spark resource slots and 
assigning \cus. During the termination of the \rp-Agent, the LRM is shutting down the Spark
cluster using Spark’s \texttt{sbin/stop-all.sh} script, which stops both the master and the slave
nodes. Similarly, the Spark specific methods for launching and managing \computeunits on Spark are 
encapsulated in a Task Spawner and Launch Method component.

\section{Experiments and Evaluation}
\label{sec:experiments}

To evaluate the \rp YARN and Spark extension, we conduct two experiments: in
Section~\ref{sec:startup_pilot_unit}, we analyze and compare \rp and \rp-YARN
with respect to startup times of both the \pilot and the \computeunits.  We use
the well-known K-Means algorithm to investigate the performance and runtime
trade-offs of a typical data-intensive application. Experiments are performed on
two different XSEDE allocated machines: Wrangler~\cite{wrangler} and
Stampede~\cite{stampede}. On Stampede every node has 16 cores and 32 GB of
memory; on Wrangler 48\,cores and 128\,GB of memory.

\subsection{Pilot Startup and \computeunit Submission}
\label{sec:startup_pilot_unit}

\begin{figure}
 \centering
 \includegraphics[width=0.45\textwidth]{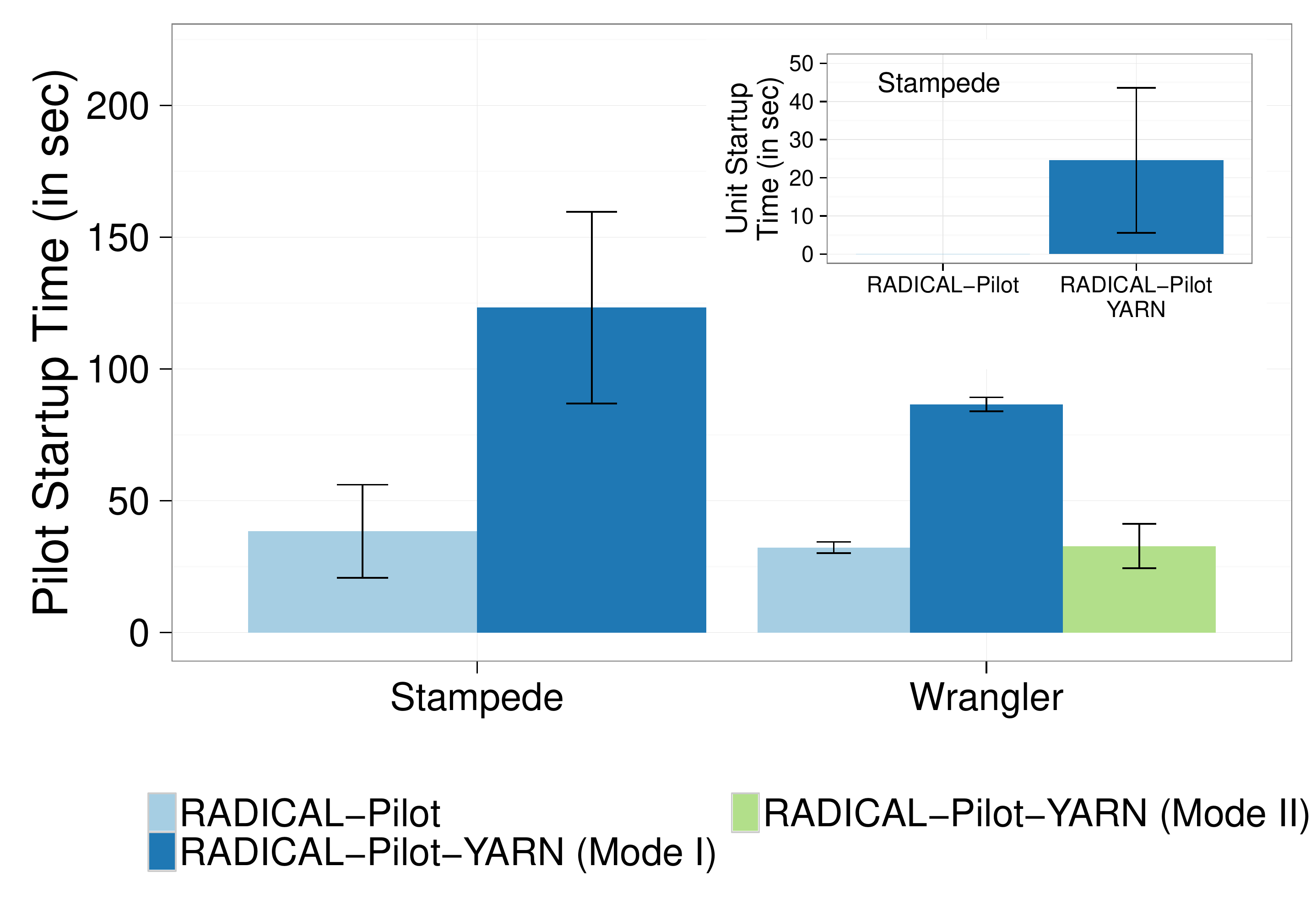}
 \caption{\textbf{\rp and \rp-YARN Overheads:} The agent startup time is higher for YARN due to 
 the overhead for spawning the YARN cluster. The inset shows that the \computeunit 
 startup time (time between application submission to YARN and startup) is also 
 significantly higher for YARN.
 \label{fig:startup_yarn}}
\end{figure}

In Figure~\ref{fig:startup_yarn} we analyze the measured overheads when starting
\rp and \rp-YARN, and when submitting \computeunits. The agent startup time for
\rp-YARN is defined as the time between \rp-Agent start and the processing of
the first \computeunit. On Wrangler, we compare both Mode I (Hadoop on HPC) and
Mode II (HPC on Hadoop). For Mode I the startup time is higher compared to the
normal \rp startup time and also compared to Mode II.  This can be explained by
the necessary steps required for download, configuration and start of the YARN
cluster. For a single node YARN environment, the overhead for Mode I (Hadoop on
HPC) is between 50-85\,sec depending upon the resource selected. The startup
times for Mode II on Wrangler -- using the dedicated Hadoop environment provided
via the data portal -- are comparable to the normal \rp startup times as it is
not necessary to spawn a Hadoop cluster.

In the inset of Figure~\ref{fig:startup_yarn} we investigate \computeunits
via \rp to a YARN cluster. For each \cu, resources have to be requested in two
stages: first the application master container is allocated followed by the
containers for the actual compute tasks. For short-running jobs this represents
a bottleneck. In the future, we will optimize this process by re-using the YARN
application master and containers, which will reduce the startup time
significantly.

In summary, while there are overheads associated with execution inside of YARN,
we believe these are acceptable, in particular for long-running tasks. The novel
capabilities of executing HPC tasks and YARN tasks within the same application
has significant benefits for which measured overheads are likely acceptable.

\subsection{K-Means}

We compare the time-to-completion of the K-Means algorithm running on two
infrastructures on HPC and HPC/YARN (Mode I Hadoop on HPC). We use three
different scenarios: 10,000 points and 5,000 clusters, 100,000 points / 500
clusters and 1,000,000 points / 50 clusters. Each point belongs to a three
dimensional space. The compute requirements is dependent upon the product of the
number of points and number of clusters, thus it is constant for all three
scenarios. The communication in the shuffling phase however, increases with the
number of points. For the purpose of this benchmark, we run 2 iterations of
K-Means. 

\begin{figure}[t]
  \centering
  \includegraphics[width=.49\textwidth]{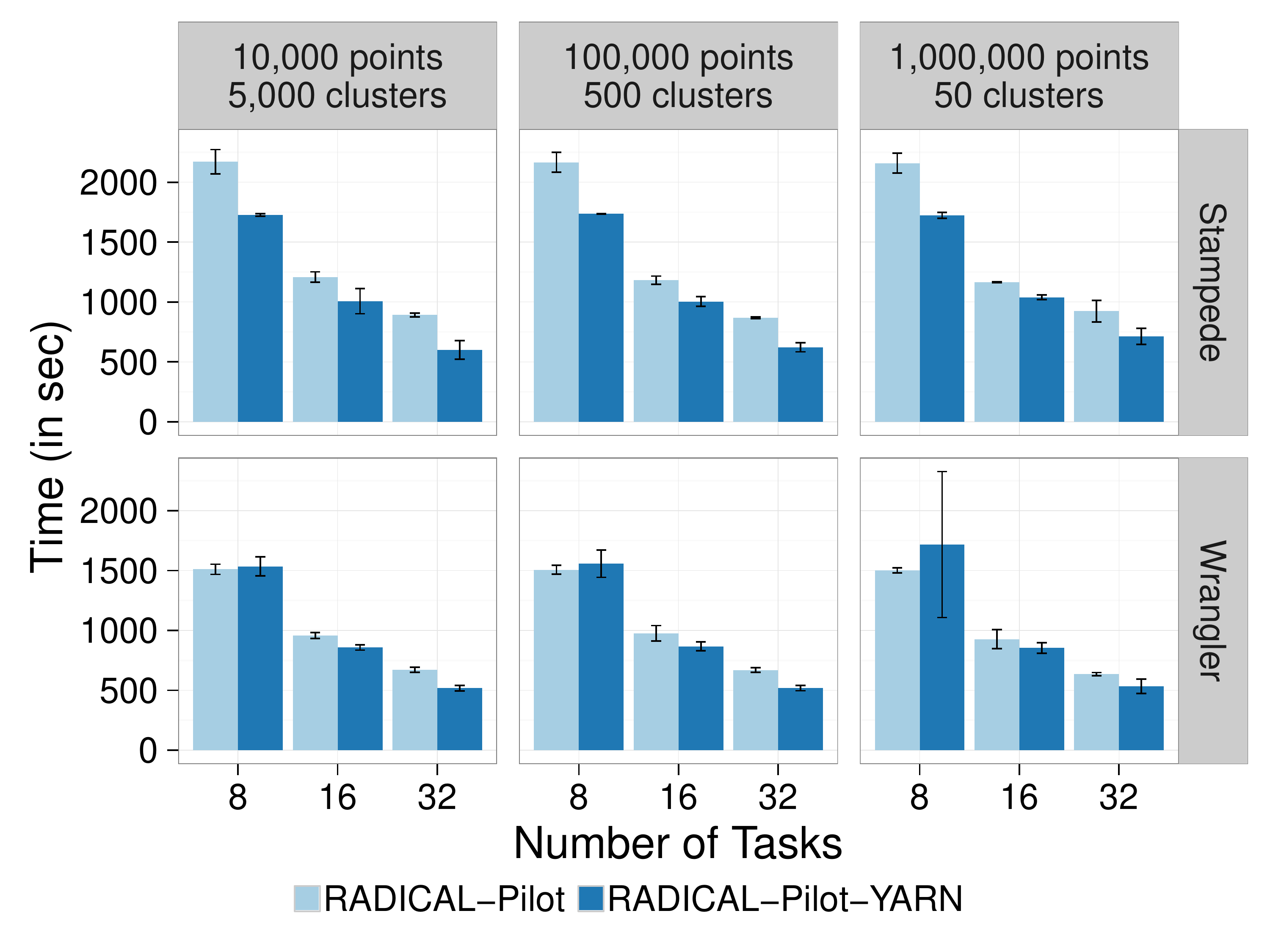}
  \caption{\rp and YARN-based K-Means on Stampede and Wrangler}
  \label{fig:experiments_kmeans_rpyarnkmeans}
\end{figure}

We utilize up to 3 nodes on Stampede and Wrangler. On Stampede every node has 16
cores and 32 GB of memory; on Wrangler 48\,cores and 128\,GB of memory. We
perform the experiments with the following configurations: 8~tasks on 1~node,
16~tasks on 2~nodes and 32~tasks on 3~nodes. For \rp-YARN, we use Mode II (Hadoop on
HPC): the YARN Resource Manager is deployed on the machine running the
\rp-Agent.  Figure~\ref{fig:experiments_kmeans_rpyarnkmeans} shows the results
of executing K-Means over different scenarios and configurations.  For \rp-YARN
the runtimes include the time required to download and start the YARN cluster on
the allocated resources.

Independent of the scenario, the runtimes decrease with the number of tasks. In
particular, for the 8 task scenarios the overhead of YARN is visible. In
particular for larger number of tasks, we observed on average 13\,\% shorter
runtimes for \rp-YARN. Also, \rp-YARN achieves better speedups, e.\,g., 3.2 for
32 tasks for the 1 million points scenario, which is significantly higher than
the \rp speedup of 2.4 (both on Wrangler and compared to base case of 8 tasks).
One of the reason for this is that for \rp-YARN the local file system is used,
while for \rp the Lustre filesystem is used.

For similar scenarios and task/resource configuration, the runtimes on Wrangler
show a significant performance improvements over Stampede. This is attributed to
the better hardware (CPUs, memory). In particular for \rp-YARN we observed on
average higher speedups on Wrangler, indicating that we saturated the 32 GB of
memory available on each Stampede node.

The amount of I/O between the map and reduce phase depends on the number of
points in the scenario. With increased I/O typically a decline of the speedup
can be observed. On Stampede the speedup is highest for the 10,000 points
scenario: average of 2.9 for \rp-YARN, and decreases to 2.4 for 1,000,000
points. Interestingly, we do not see the effect on Wrangler indicating that we
were not able to saturate the I/O system with our application.

In summary, despite the overheads of \rp-YARN with respect to \pilot and
\computeunit startup time, we were able to observe performance improvements (o
average 13\,\% better time to solution) mainly due to the better performance of
the local disks.

\section{Discussion and Conclusion}
\label{sec:conclusion}

{\bf Discussion:} Integrating \yarn with existing HPC platforms will enable a
range of applications to take advantage of the advances in the Hadoop ecosystem
that are not possible currently. The \pilot-based approach provides a common
framework for both HPC and YARN application over dynamic resources. The
challenges associated with this range from the conceptual to the practical.

One of the practical considerations when integrating Hadoop and HPC arises from
the fact that Hadoop is typically deployed as system-level framework and
provides many configurations options. For achieving optimal performance Hadoop
configurations should be fine-tuned so as to optimally exploit memory, cores,
SSDs, parallel filesystems and external libraries (e.\,g., for high-performance
linear algebra). Currently, SAGA-Hadoop and \rp are able to detect and optimize
Hadoop with respect to memory and core usage. In the future, we will provide
configuration templates so that resource specific hardware can be exploited,
e.\,g.\, available SSDs can significantly enhance the shuffle performance.

Several methods for optimizing Hadoop performance on HPC have been
proposed. Most of them address the performance sensitive shuffle phase, which is
characterized by significant data movements. Intel provides a Hadoop Lustre
adaptor~\cite{lustre_hadoop} that optimizes data movements in the shuffle phase
by replacing the built-in MapReduce shuffle implementation. Panda et al.
propose the usage of high-performance network capabilities, e.\,g., RDMA
for optimizing the shuffle phase in MapReduce~\cite{homr} and
Spark~\cite{rdma_spark} by replacing the existing Java socket based
communication.

Another challenge is the wide spectrum of programming models that can be found
in the Hadoop ecosystem and in HPC. Combining programming models and runtimes,
such MapReduce, Spark, Flink with HPC approaches, such as OpenMP, MPI and
GPU/Cuda approaches is difficult and subject to different constraints and
optimizations: (i) Should the frameworks be used side-by-side and data moved
between them? (ii) Should HPC sub-routines be called from a Hadoop framework?,
or (iii) Should Hadoop be called from HPC? Many hybrid approaches for HPC and
Hadoop integration exploring options (i) and (ii) have been proposed: \rp
enables HPC and Hadoop application to execute {\it side-by-side} and supports
data movements between these environments. Different Hadoop and Spark frameworks
re-use HPC code, e.\,g.\, MLlib~\cite{mllib} relies on HPC BLAS libraries and
SparkNet~\cite{2015arXiv151106051M} utilizes the GPU code from Caffe. Typically,
these native codes are called by sub-routines in the map or reduce phase either
via command line, JNI or other language bridges. While it is possible to call
arbitrary code from Spark or other Hadoop frameworks, integrating codes that
explore parallelism outside of Spark is not straightforward and may lead to
unexpected results as the Spark scheduler is not aware of these. While one can
force certain resource configurations, typically the necessary tricks do not
generalize to other resources/infrastructures and violate the Spark programming
and execution model.

Another similar question is: which infrastructure and programming model should
be used to develop a new applications? When is it worth to utilize hybrid
approaches? While data filtering and processing is best done with Hadoop or
Spark (using e.\,g., the MapReduce abstraction), the compute-centric parts of
scientific workflows are best supported by HPC. Utilizing hybrid environments is
associated with some overhead, most importantly data needs to be moved, which
involves persisting files and re-reading them into Spark or another Hadoop
execution framework. In the future it can be expected that data can be directly
streamed between these two environments; currently such capabilities typically
do not exist. Application-level scheduling approaches, e.\,g.\, using the
\pilot-Abstraction, enable applications to reason about these trade-offs, as
scheduling decisions are highly application dependent taking into account data
locality, necessary data movements, available resources and frameworks.

Infrastructure are becoming more heterogeneous: multicore, accelerators, more
levels of memory (e.\,g., non-volatile memory) will increase the deployment
challenges. New HPC resources, such as Wrangler attempt to balance the diverse
requirements by offering HPC and dedicated Hadoop environments leaving the
choice to the user. At the same time, software heterogeneity, size and
complexity continues increase as tools start to explore their new infrastructure
capabilities.  There are several investigations with respect to the usage of GPU
inside of Spark, e.\,g.\ for deep learning~\cite{2015arXiv151106051M}. While
convergence is desirable, there is a long road ahead of us.

{\bf Conclusion:} Hadoop and Spark are used by increasing number of scientific
applications mainly due to accessible abstractions they provide. HPC and Hadoop
environments are converging and cross-pollinating, for example, as shown by
MLLib/Spark utilization of linear algebra BLAS library that originated in HPC,
increasingly parallel and in-memory computing concepts that originated in HPC
are adopted by Hadoop frameworks. As this trend strengthens, there will be a
need to integrate HPC and Hadoop environments, e.\,g.\ for combining multi-stage
scientific applications comprising of a simulation and analytics stage.
Currently, traditional HPC applications lack the ability to access and use
Hadoop and other Hadoop tools, without sacrificing the advantages of HPC
environments. One prominent reason behind the limited uptake, is related to
finding satisfactory and scalable resource management techniques usable for
Hadoop frameworks on HPC infrastructure.

This paper motivates the use of the \pilot-Abstraction as an integrating
concept, discusses the design and implementation of \rp extensions for Hadoop
and Spark, and validates them with scalability analysis on two supercomputers: a
traditional Beowulf-style mammoth machine (Stampede), and a special purpose
data-intensive supercomputer (Wrangler). Our experiments use \rp, and introduce
two extensions to \rp to better integrate Hadoop on HPC. We demonstrate that the
\pilot-Abstraction strengthens the state of practice in utilizing HPC resources
in conjunction with emerging Hadoop frameworks by allowing the user to combine a
diverse set of best-of-breed tools running on heterogeneous infrastructure
consisting of HPC. Using these capabilities, complex data applications utilizing
a diverse set of Hadoop and HPC frameworks can be composed enabling scalable
data ingests, feature engineering \& extractions and analytics stages. Each of
these steps has its own I/O, memory and CPU characteristics. Providing both a
unifying and powerful abstraction that enables all parts of such a data pipeline
to co-exist is critical.

\emph{Future Work:} This work provides a starting point for multiple lines of
research and development. We have begun work with biophysical and molecular
scientists to integrate molecular dynamics data trajectory analysis
capabilities. These include principal component based
analysis~\cite{doi:10.1021/ct400341p}, as well as graph-based
algorithms~\cite{mdanalysis}. While we demonstrate the importance of integrated
capabilities, we will also extend the \pilot-Abstraction to support improved
scheduling, e.\,g.\, by improving the data-awareness, introducing predictive
scheduling and other optimization. We will evaluate support for further
operations, e.\,g.\, to execute collectives on data, utilizing in-memory
filesystems and runtimes (e.\,g., Tachyon and Spark) for iterative algorithms.
Also, infrastructures are evolving: container-based virtualization (based on
Docker~\cite{docker}) is increasingly used in cloud environments and also
supported by \yarn. Support for these emerging infrastructures is being added to
the \pilot-Abstraction.

\noindent{\textbf {Author Contribution: }AL implemented SAGA-Hadoop and initial prototypes for
  \pilot-Hadoop and \pilot-Spark, and initial experiments, as well as the majority
  of the writing. IP and GC implemented second generation prototypes for
  \pilot-Hadoop and \pilot-Spark, as well as associated experiments. IP also
  provided the related work section. AL and SJ designed experiments and analysis
  was performed by all.}

\noindent\textbf{Acknowledgement} We acknowledge Pradeep Mantha for early experiments.
We acknowledge financial support of NSF 1440677 and NSF 1443054. This work has
also been made possible thanks to computer resources provided by XRAC award
TG-MCB090174 and an Amazon Computing Award. We thank Professor Geoffrey Fox for
useful discussions.

\end{document}